\documentclass[aps,pre,showpacs,preprintnumbers,twocolumn,nofootinbib,superscriptaddress]{revtex4-1}

\usepackage{geometry}                % See geometry.pdf to learn the layout options. There are lots.
\usepackage{graphicx}
\usepackage{amssymb}
\usepackage{dsfont}
\usepackage{epstopdf}
\usepackage{color}
\usepackage{mathtools}
\usepackage{cancel}
\usepackage{amsmath}

\definecolor{red}{rgb}{1,0,0}
\definecolor{blue}{rgb}{0,0,1}
\definecolor{black}{rgb}{0,0,0}

\newcommand{\p}{\partial}
\newcommand{\eq}[1]{\begin{align}#1\end{align}}
\newcommand{\eqs}[1]{\begin{align*}#1\end{align*}}
\newcommand{\ffrac}[2]{\mbox{$\frac{#1}{#2}$}}

\newcommand{\TT}{\mathcal{T}}

\newcommand{\LL}{{\mathcal{L}}}

\newcommand{\D}{{\mathcal{D}}}

\usepackage{calrsfs}

\usepackage{scalerel,stackengine}
\stackMath
\newcommand\widecheck[1]{%
\savestack{\tmpbox}{\stretchto{%
  \scaleto{%
    \scalerel*[\widthof{\ensuremath{#1}}]{\kern-.6pt\bigwedge\kern-.6pt}%
    {\rule[-\textheight/2]{1ex}{\textheight}}%WIDTH-LIMITED BIG WEDGE
  }{\textheight}% 
}{0.5ex}}%
\stackon[1pt]{#1}{\scalebox{-1}{\tmpbox}}%
}

\newcommand{\PP}{\mathbb{P}}

\begin{document}
\title{Robustness of the Random Language Model}

\author{Fatemeh Lalegani}
\author{Eric De Giuli}
\affiliation{Department of Physics, Toronto Metropolitan University, M5B 2K3, Toronto, Canada}
\date{\today}

\newcommand{\Ld}{{L^{\dagger}}}
\newcommand{\Rd}{{R^{\dagger}}}
\newcommand{\Qd}{{Q^{\dagger}}}
\newcommand{\Pd}{{P^{\dagger}}}
\newcommand{\Lamd}{{\Lambda^{\dagger}}}

\begin{abstract}
The Random Language Model (De Giuli 2019) \cite{DeGiuli19} is an ensemble of stochastic context-free grammars, quantifying the syntax of human and computer languages. The model suggests a simple picture of first language learning as a type of annealing in the vast space of potential languages. In its simplest formulation, it implies a single continuous transition to grammatical syntax, at which the symmetry among potential words and categories is spontaneously broken. Here this picture is scrutinized by considering its robustness against { extensions of the original model, and trajectories through parameter space different from those originally considered. It is shown here that (i) the scenario is robust to explicit symmetry breaking, an inevitable component of learning in the real world; and (ii) the transition to grammatical syntax can be encountered by fixing the deep (hidden) structure while varying the surface (observable) properties. It is also argued that the transition becomes a sharp thermodynamic transition in an idealized limit. Moreover, comparison with human data on the clustering coefficient of syntax networks suggests that the observed transition is equivalent to that normally experienced by children at age 24 months. The results are discussed in light of theory of first-language acquisition in linguistics, and recent successes in machine learning.}

%This single-transition scenario is considered in the light of theory in linguistics.
%inclusion of inevitable complications in the real world
%elaborated upon, both from the perspective of a learning algorithm, and 

\end{abstract}

\maketitle

Language is a way to convey complex ideas, instructions, and structures through sequences. While ubiquitous in everyday life, it also has a central role in computer science and molecular biology. One can ask if these disparate applications of language have any common features. The answer, apparently, is positive: the formalism of generative grammar, due to Post and Chomsky \cite{Post43,Chomsky02}, though initially developed for human language, was immediately applied to computer languages, where it has remained important \cite{Hopcroft07}, and has also been applied to the molecular languages spoken by the cell \cite{Searls02,Knudsen03}. Other idiosyncratic applications highlight the flexibility of the approach \cite{Escudero97}. 

Generative grammar models the syntax of language by a set of rules that, upon repeated application, yield `grammatical'  sentences. In this framework, for any grammatical sentence, there is a latent `derivation' structure that encodes the syntax of that sentence; some examples are shown in Fig. 1. 

{ In the computer science and linguistics literature, research on generative grammars focuses on classifications and algorithms: classifications of grammars based on the complexity of the rules, corresponding classifications on the types of simple computers (automata) that can read languages, and algorithms to parse text. Many results exist on the time and resource cost of parsing \cite{Hopcroft07}. Yet, if we admit that languages are always used by systems embedded in the physical world, then new questions arise: how much energy does it take to parse a grammar of a given complexity \cite{Wolpert23}? How does a child navigate the space of all potential languages to hone in on the one taught to her? More broadly, one can ask, in the spirit of statistical physics, whether large grammars will show universality of the same type familiar from equilibrium statistical mechanics.

}
As an inroad to these questions, in Ref.\cite{DeGiuli19} the senior author proposed a simple ensemble of context-free grammars, the class of grammars most relevant to human and computer language. In its stochastic version, a context-free grammar assigns a probability (or more generally a weight) to each rule. Ref.\cite{DeGiuli19} explored the information-theoretic properties of grammars as functions of the variance of rule weights, the number of hidden categories, and the number of words.  

The main result of Ref.\cite{DeGiuli19} is that the entropy of text produced by a context-free grammar depends strongly on the variance of the weights, such that two regimes are seen: a simple one in which, despite the presence of stochastic rules, sentences are nearly indistinguishable from uniform random noise; and a complex one in which sentences convey information. The transition between these regimes could be understood as a competition between Boltzmann entropy and an energy-like quantity.

\begin{figure}[t!]
\includegraphics[width=\columnwidth]{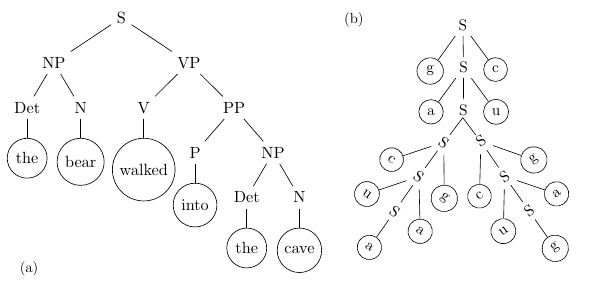}
\caption{ Illustrative derivation trees for (a) simple English sentence, and (b) RNA secondary structure (after \cite{Searls02}). The latter is a derivation of the sequence `gacuaagcugaguc' and shows its folded structure. Terminal symbols are encircled. Figure reproduced from \cite{DeGiuli19}.
}\label{fig1}
\end{figure}

This work left many questions open: 

(i) is the schematic learning scenario of  Ref.\cite{DeGiuli19} robust to inevitable complications of real-world human language learning, such as explicit symmetry breaking?

(ii) is the transition shown in Ref.\cite{DeGiuli19} a true thermodynamic phase transition in an appropriate thermodynamic limit?

(iii) can the RLM be solved analytically?

(iv) what are the energy costs of physical systems that use CFGs to produce text?

In this work we address (i) and (ii) and comment on (iii); (iv) will be addressed elsewhere. We first show how previous theory implies that the RLM transition can be reached by increasing the heterogeneity of surface rules, and confirm this numerically. Then we consider the learning problem and motivate the RLM with a bias. Simulating this, we see that the RLM transition is preserved, but shifted due to the bias. A simple theory can rationalize the initial onset of nontrivial sentence entropy. To compare with human data in Ref.\cite{Corominas-Murtra09} we measure the clustering coefficient of a sentence graph, constructed from sampled sentences. This clustering is small until the RLM transition, where it begins to grow. Such a growth in clustering is also observed in syntactic networks made from human data, and supports that the RLM transition is equivalent to that typically experienced by children around 24 months. Finally, we discuss these results in the light of linguistic theory on first language acquisition.

\section{Brief review of the Random Language Model}

To establish notation, here we briefly review the RLM. Without loss of generality CFGs are assumed to be in Chomsky normal form, so that rules either take one hidden symbol $a$ to two hidden symbols $b,c$, or one hidden symbol $a$ to an observable one, $B$. These are quantified by weights $M_{abc}$ and $O_{aB}$, respectively. For a sentence $o_j, j = 1, \ldots, \ell$ with derivation $\sigma_j, j = 1, \ldots, 2\ell-1$ on the tree $\mathcal{T}$, define $\pi_{abc}(\sigma)$ as the (unnormalized) usage frequency of rule $a\to bc$ and $\rho_{aB}(\sigma,o)$ as the (unnormalized) usage frequency of $a\to B$. { Let the number of hidden symbols be $N$, and the number of observable symbols be $T$. }Then consider the energy function
\eq{ \label{E}
E(\sigma,o; M,O) = -\sum_{a,b,c} \pi_{abc} \log M_{abc}  - \sum_{a,B} \rho_{aB} \log O_{aB} .
}
The Boltzmann weight $e^{-\beta E}$ counts derivations with a multiplicative weight $(M_{abc})^\beta$ for each usage of the interior rule $a \to bc$, and weight $(O_{aB})^\beta$ for each usage of the surface rule $a \to B$. We furthermore assign a weight to the tree itself: if each hidden node gets a weight $q$ and each surface node gets a weight $p$, then a rooted tree with $\ell$ leaves gets a weight $q^{\ell-1} p^\ell$. The relative probability $p/q$ controls the size of trees; as in Ref.\cite{DeGiuli19} we fix $q=1-p$ and set $p=1/2+\epsilon$ where $\epsilon\ll 1$ to get large trees. 

{ Given the grammar, the probability of a derivation is then
\eq{
\PP(\mathcal{T}, \sigma,o | M,O) = \frac{1}{Z} q^{\ell-1} p^\ell e^{-\beta E}
}

Note that although we write the weight of a derivation in a Boltzmann-like form, the actual form of the weight is simply the conventional definition of a stochastic context-free grammar. }

The RLM is an ensemble of CFGs. In Ref.\cite{DeGiuli19} it was argued that a generic model will have lognormally distributed weights, { so that the probability of a grammar is }
\eq{ \label{PPG}
\PP_G(M,O) & \equiv Z_G^{-1} \; J \; e^{-\epsilon_d s_d}  e^{- \epsilon_s s_s } 
}
where $s_d$ and $s_s$ are defined by 
\eq{ \label{s1}
s_d = \frac{1}{N^3} \sum_{a,b,c} \log^2 \left[\frac{M_{abc}}{\overline{M}}\right], \;\; s_s =  \frac{1}{NT} \sum_{a,B} \log^2\left[\frac{O_{aB}}{\overline{O}} \right]
%s_d = \frac{1}{N^3} \sum_{a,b,c} \left|\log \frac{M_{abc}}{\overline{M}} \right|^2, \quad s_s =  \frac{1}{NT} \sum_{a,B} \left|\log \frac{O_{aB}}{\overline{O}} \right|^2
}
and $J = e^{-\sum_{a,b,c} \log M_{abc} - \sum_{a,B} \log O_{aB}}$. Here $\overline{M}=1/N^2$ and $\overline{O}=1/T$. %A plot of the weights for a range of $\tilde\epsilon_d=\epsilon_d/N^3$ is shown in Fig.\ref{figens}a. 
It is straightforward to show that $\epsilon_d$ and $\epsilon_s$ satisfy
\eq{ \label{s2}
\overline{s_d} = (2\tilde \epsilon_d)^{-1}, \qquad \overline{s_s} = (2\tilde \epsilon_s)^{-1}.
}
where $\overline{\;\cdot\;}$ denotes a grammar average and $\tilde \epsilon_d = \epsilon_d/N^3$, $\tilde \epsilon_s = \epsilon_s/(NT)$. 

{ Two arguments were given in Ref.\cite{DeGiuli19} for the lognormal distribution: first, since languages must be comprehensible to a variety of speakers at any moment, they cannot evolve rapidly. If they evolve slowly under independent multiplicative adjustments to the weights, then a lognormal distribution follows by the multiplicative version of the central limit theorem \cite{Sornette97}. Indeed the lognormal distribution is ubiquitous for the distributions of positive random variables, such as transition weights, in real-world systems \cite{Broido19}. In this interpretation, $\epsilon_d$ and $\epsilon_s$ are general control parameters for the ensemble.

A second independent argument is to assume that $s_d$ and $s_s$ are the relevant quantities to characterize grammars in the course of learning; then a lognormal follows by a maximum entropy argument. The quantities $s_d$ and $s_s$ could be motivated {\it a priori} as appropriate measures of heterogeneity, or {\it a posteriori} by the observation that they control the Shannon entropy of sequences (along with $N$ and $T$). In this interpretation, $\epsilon_d$ and $\epsilon_s$ are Lagrange multipliers that enforce the expected values of $s_d$ and $s_s$. }

\begin{figure}[th!]
\centering
%\begin{subfigure}[b]{0.49\textwidth}\centering
\includegraphics[width=.95\columnwidth]{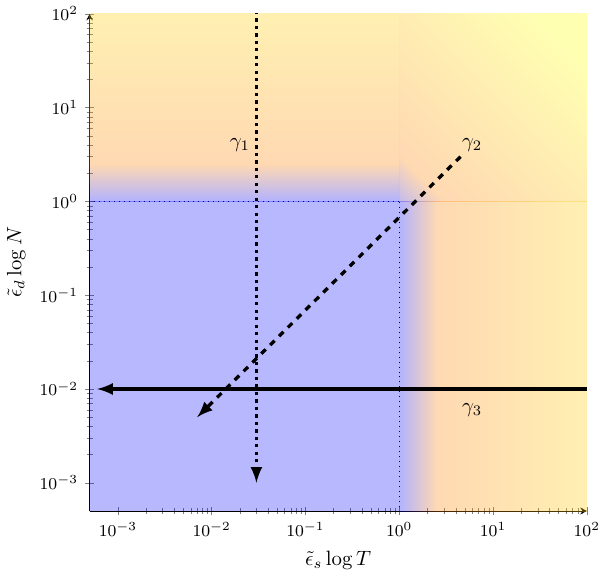}
\caption{Phase diagram of the RLM, in the replica-symmetric approximation. Text is grammatical in the lower-left region, demarcated approximately by  $\tilde\epsilon_s \log T\approx 1, \tilde\epsilon_d \log N\approx 1$ (light dotted). Three paths $\gamma_j$ through the diagram are sketched: $\gamma_1$ at fixed $\tilde\epsilon_s$, considered in \cite{DeGiuli19}; $\gamma_2$ with $\tilde\epsilon_s=\tilde\epsilon_d$, discussed below; and $\gamma_3$ at fixed $\tilde\epsilon_d$, also discussed below.
}\label{fig_phase}
\end{figure}

Let us show how $\beta$ can be scaled out of the problem. Consider the grammar and derivation average of a generic observable of a derivation $\mathcal{O}(\sigma,o)$:
\eq{ 
\overline{ \mathcal{O} } & =  \ffrac{1}{Z Z_G} \int dM dO \; J(M,O) \; e^{-\epsilon_d s_d(M)}  e^{- \epsilon_s s_s(O) } \notag\\
& \qquad \times \sum_{\sigma,o} e^{-\beta E(\sigma,o; M,O) } \mathcal{O}(\sigma,o)
}
Making a change of variable $M_{abc}^\beta = M'_{abc}$,  $O_{aB}^\beta = O'_{aB}$ we get
\eq{
\overline{ \mathcal{O} } & =  \ffrac{1}{\beta^{N^3+NT}}\ffrac{1}{Z Z_G} \int dM' dO' \; J(M',O') \; e^{-\frac{\epsilon_d}{\beta^2} s_d'(M')} \notag\\
&\qquad \times e^{- \frac{\epsilon_s}{\beta^2} s_s'(O') }\sum_{\sigma,o}  e^{- E(\sigma,o; M,O) } \mathcal{O}(\sigma,o) ,
}
where $s_d'(M), s_s'(O)$ are defined as in \eqref{s1} with the replacement $\overline{M} \to \overline{M}^\beta$, $\overline{O} \to \overline{O}^\beta$. It follows that the parameters $\epsilon_d$, $\epsilon_s$, and $\beta$ do not affect observables independently, but only in the ratios $\epsilon_d/\beta^2$ and $\epsilon_s/\beta^2$, up to the other trivial modifications. In particular, increasing temperature is equivalent to increasing $\epsilon_d$ and $\epsilon_s$. For this reason, in Ref.\cite{DeGiuli19} these parameters were called deep and surface temperatures, respectively. From now on we set $\beta=1$.

%A small `deep temperature' $\epsilon_d$ corresponds to a large deep sparsity.
The model \eqref{PPG} was called in Ref.\cite{DeGiuli19}  the Random Language Model (RLM). The properties of the sentences as a function of grammar heterogeneity were studied in Refs.\cite{DeGiuli19,DeGiuli19a,De-Giuli22}. The main result of Ref.\cite{DeGiuli19} is that as $\epsilon_d$ is lowered, there is a transition between two regimes at $\epsilon_d \approx N^3 \log^\alpha N$ where $\alpha=1$ or $\alpha=2$ depending on the quantity considered. Theory in Refs.\cite{DeGiuli19a,De-Giuli22} predicts this scaling (with $\alpha=1$) and also predicts that the transition can be reached by fixing $\epsilon_d$ but lowering $\epsilon_s$. 

Theory for the RLM was developed in Refs.\cite{DeGiuli19a,De-Giuli22}, with final results obtained in the replica-symmetric approximation. For a text of $m$ sentences and total length $\ell$, the result of Refs.\cite{DeGiuli19a,De-Giuli22} is that the Boltzmann entropy of configurations is
\eq{ \label{SRS1}
S_{RS} & = (\ell-m) \log (g N^2/h) + \ell \log (g T h) \notag\\
& \qquad - \frac{\ell}{4\tilde\epsilon_s} - \frac{\ell-m}{4\tilde\epsilon_d} + S_{\ell,m} ,
}
where $S_{\ell,m}$ is a combinatorial coefficient independent of the other parameters, and $g$ and $h$ are couplings that control the size of trees. In the considered limit of large trees $g= h \approx 1/\sqrt{8}$.

\begin{figure*}[th!]
\centering
%\begin{subfigure}[b]{0.49\textwidth}\centering
\includegraphics[width=.7\textwidth]{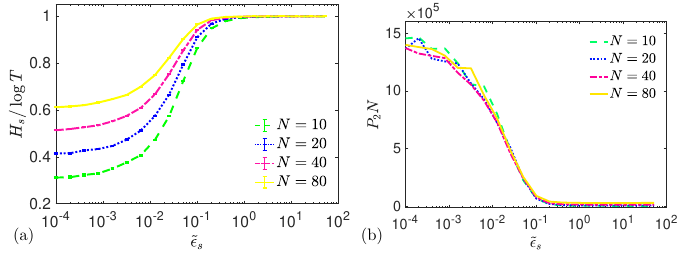}
\caption{The RLM transition can be encountered by lowering the surface temperature $\epsilon_s$. Curves are shown at $T=1000$, $\tilde\epsilon_d \approx 0.03$, and indicated values of $N$; (a) the surface entropy drops around $\tilde\epsilon_s\approx 1/\log T$, while (b) the surface order parameter $P_2$ increases as $\tilde\epsilon_s$ is lowered.  }
\label{fig2}
\end{figure*}

Now, by a standard argument \cite{Parisi88} the Boltzmann entropy of configurations is equal to the Shannon entropy of the probability distribution over configurations. This latter quantity can be written as the entropy of forests at given $m$ and $\ell$, plus the conditional entropy of hidden configurations on those trees, plus the conditional entropy of leaves on those configurations. Each of these entropies can be written as the corresponding rate multiplied by the number of symbols. There are $\ell$ observable symbols and $2\ell-m$ hidden symbols, but all roots are set to the start symbol. Thus
\eq{ \label{SRS2}
S_{RS} = S_{\text{forest}}(\ell,m) + (2\ell-2m) H_d + \ell H_{s|d} ,
}
where $H_d$ is the entropy rate of hidden symbols and $H_{s|d}$ is the conditional entropy rate of observable symbols, given the hidden ones. These configurational entropies are trivial at $\tilde\epsilon_d,\tilde\epsilon_s=\infty$ so that we can write
\eqs{
& S_{RS}(\tilde\epsilon_d =\infty,\tilde\epsilon_s=\infty) \\
& = (\ell-m) \log (g N^2/h) + \ell \log (g T h) + S_{\ell,m}  \\
& = S_{\text{forest}}(\ell,m) + (2\ell-2m) \log N + \ell \log T
}
The factors of $\log N$ and $\log T$ cancel from this equality, as they must. As a result we obtain $S_{\text{forest}}(\ell,m)$ and finally
\eq{ \label{SRS3}
S_{RS} & = S_{\text{forest}}(\ell,m)+  (\ell-m) \log (N^2) + \ell \log (T) \notag\\
& \qquad - \frac{\ell}{4\tilde\epsilon_s} - \frac{\ell-m}{4\tilde\epsilon_d} .
}
Comparing with \eqref{SRS2} and noting that this equation must hold for all $\ell,m$ (with $\ell,m\to \infty, \ell/m$ finite ), we deduce
\eq{ \label{Hd}
H_d = \log N -  \frac{1}{8\tilde\epsilon_d} , \qquad H_{s|d} = \log T - \frac{1}{4\tilde\epsilon_s} ,
}
in the replica-symmetric approximation. Since these entropies cannot be negative, they give lower bounds on the validity of the replica-symmetric approximation (in $\tilde\epsilon_d,\tilde\epsilon_s$ space). At small enough $\tilde\epsilon_d$ or $\tilde\epsilon_s$, the approximations used to derive \eqref{Hd} must break down. It also follows from this that the normalized entropies $H_d/\log N$ and $H_{s|d}/\log T$ should collapse with $\tilde\epsilon_d \log N$ and $\tilde\epsilon_s\log T$, respectively.

Note that Ref.\cite{DeGiuli19} measured $H_s$, not $H_{s|d}$. In general the Bayes rule for conditional entropy is $H_{s|d} = H_s - H_d + H_{d|s}$. When $\tilde\epsilon_s$ is small, then knowing the observable symbols also fixes their POS tags, so $H_{d|s}\approx 0$ and $H_s(\tilde\epsilon_s\ll 1) \approx H_s-H_d$. However when $\tilde\epsilon_s$ is large, then knowing the hidden symbol tells you nothing about the observable symbol, so $H_{s|d}\approx H_s$.  Thus generally we expect that as a function of $\tilde\epsilon_s$, $H_s$ behaves similarly to $H_{s|d}$. 

{ We emphasize that although $\epsilon_d$ and $\epsilon_s$ play parallel roles in the distribution, and in many aspects of theory, they are distinct parameters with asymmetric control over observables, since the hidden structure of trees affects sentences but not vice-versa. Roughly speaking, we can demarcate four regimes. To explain these, we use the example of phrase structure, where observable symbols are words and hidden symbols are abstract categories, like noun phrase (NP), verb phrase (VP), verb (V), and so on. In syntax trees, the hidden symbols that appear just above the leaves are called part-of-speech (POS) tags -- symbols like verb, noun, adjective, and so on. 

If $\tilde\epsilon_d \log N \gg 1$ while $\tilde\epsilon_s \log T \ll 1$, then sentences will consistently match words with their POS tags, but there will be no syntactic structure connecting the words together. Conversely if  $\tilde\epsilon_d \log N \ll 1$ while $\tilde\epsilon_s \log T \gg 1$, then sentences have structure, but the final observable words are randomly assigned from POS categories. If both of these parameter combinations are large, then sentences lack all structure; while if both are small, then sentence structure is complete. This phase diagram is sketched in Fig.\ref{fig_phase}, along with 3 paths through the space.

As a consequence, one can discuss learning by different routes through ($\tilde\epsilon_d,\tilde\epsilon_s)$ space (in addition to variations in $N$ and $T$ which could also be considered). In particular, theory predicts that the RLM transition can be probed by fixing $\tilde\epsilon_d$ and lowering $\tilde\epsilon_s$. We now show that this prediction is verified by numerics. }

\subsection{The RLM transition is encountered by increasing surface heterogeneity}

We simulated the RLM with $T=1000$ and $\tilde\epsilon_d \approx 0.03$ at various values of $N$ and $\tilde \epsilon_s$. For each parameter value, 60 distinct grammars were constructed, and 200 sentences were sampled for each grammar. The results for the surface entropy are shown in Fig.\ref{fig2}; as predicted by theory, the entropy begins to drop from its trivial value at $\tilde\epsilon_s \approx 1/\log T \approx 0.1$. %On linear axes, its effect is noticeable by $\tilde\epsilon_s \approx 0.1$.

Since $\tilde\epsilon_d$ is fixed as $\tilde\epsilon_s$ varies, there is no variation in the hidden parts of the derivations: the quantities shown in Ref.\cite{DeGiuli19} to quantify the RLM transition, like the deep entropy $H_d$ and the order parameter $Q_2$, are flat as $\tilde\epsilon_s$ varies. Instead the transition can be quantified by the surface analog of the order parameter $Q_2$. For a surface rule $a \to B$ define 
\eq{
P_{aB}(M,O) = \langle \delta_{\sigma_{\alpha},a} (T \delta_{o_\alpha,B}-1) \rangle ,
}
averaged over all surface vertices $\alpha$ and over all derivations. Here $\sigma_\alpha$ is the hidden symbol and $o_\alpha$ the observable one. $P$ measures how the application of this rule differs from a uniform distrbution. An Edwards-Anderson type order parameter for surface structure is
\eq{
P_2 = \sum_{a,B} \overline{P_{aB}^2} ,
}
where $\overline{\;\;}$ is an average over grammars. This quantity is shown in Fig.\ref{fig2}b. As expected, $P_2$ increases from a small value at high $\tilde\epsilon_s$ around the transition point. 

%\subsection{The RLM transition is encountered by increasing surface heterogeneity}

\section{Learning a context-free grammar} 

Now we consider the learning problem. How does a child actually learn the specific grammar of its environment? 

Our goal is not to completely answer this question, but simply to motivate why and how the symmetry of symbols should be explicitly broken. As a simple model, we suppose that the speaker utters sentences by drawing them from a stochastic grammar, which we take to be context-free. %The grammar models syntax. 
In a stochastic grammar, the weights quantify their frequency of use, which, for learners, is a proxy for their correctness. When all the weights are equal, nothing is known, and the grammar samples uniform noise (`babbling'). In contrast, when the weights have a wide distribution, the grammar is highly restrictive and the output sentences are highly non-random. 

The learning scenario suggested in Ref.\cite{DeGiuli19} was quite generic: suppose the child knows, possibly due to physiological constraints, that she is learning a CFG. Initially she knows nothing of weights, so she starts at $\epsilon_d=\epsilon_s=\infty$. Her initial speech will be uniform random noise. Now, as she tries to mimic her caregivers, we assume that she tunes the grammar weights. In doing so the corresponding values of $\epsilon_s$ and $\epsilon_d$, which could be defined from \eqref{s2}, will inevitably decrease. Then, the prediction of the RLM is that the entropy of her speech will remain high for some time, until quite suddenly it begins to decrease. At this point her speech begins to convey information. 

This scenario is quite schematic. Let us try to make it more concrete.

Consider first an optimal learning scenario. She hears sentences $\gamma$, with words $o_j^\gamma, j=1, \ldots, \ell_\gamma$, and wants to find the optimal grammar that produces them. It is natural to maximize the log-likelihood of the grammar, given the data, given by
\eq{
\LL(M,O; o) = \log \PP( o | M, O) ,
}
which is considered as a function of the grammar, with fixed sentences $\{o\}$. { We assume that the space of grammars that she searches is the full set of possibilities, but of course physiological constraints may also play a role. } The sentence probability is 
\eq{
\PP( o | M, O) & = \prod_\gamma \sum_{\TT_\gamma,\sigma^\gamma_k} \PP( o^\gamma, \sigma^\gamma, \TT_\gamma | M, O) \\
& = \prod_\gamma \frac{1}{Z} \underbrace{\sum_{\TT_\gamma,\sigma^\gamma_k} e^{-E(o^\gamma, \sigma^\gamma, \TT_\gamma ; M, O)}}_{\equiv Z(o^\gamma)} ,
}
where $Z(o^\gamma)$ is then a partition function restricted to the given sentence $o^\gamma$. In principle, she can estimate these quantities by speaking: every sentence she speaks adds a contribution to the denominator $Z$. If she feels that her caregiver understood it, then she also adds a contribution to the numerator $Z(o^\gamma)$. 

Unfortunately computing these restricted partition functions is difficult, both analytically, and for the child. %In principle, the child can estimate 
So we consider a simpler, more idealized scenario. She keeps track of a lexicon 
\eqs{
\{ tree, mama, toy, book, open, close, eat, sleep, up, down, \ldots \} ,
}
how many times she's heard each word, and also the categories to which each word belongs 
\eqs{
\{ noun, verb, adjective, \ldots \} ,
}
called part-of-speech (POS) tags. %Thus she obtains an estimate $\tilde\rho_{aB}$ of the surface grammar weights $\rho_{aB}$. 

%She thus obtains an estimate of the word probability $\tilde p$. Then she maximizes the likelihood of $\tilde p$, 
%\eq{
%\LL(M,O; \tilde p) & = \log \PP( \tilde p | M, O) \\
%& = \log \prod_\gamma \sum_{\TT_\gamma,\sigma^\gamma_k, \o^\gamma_B} \delta_{C(o), \ell \tilde p} \PP( o^\gamma, \sigma^\gamma, \TT_\gamma | M, O) , \label{L2}
%}
%where $C(o)_B$ is the count of word $B$ in the text of total length $\ell$, i.e.
%\eq{
%C(o)_B = \sum_{j=1}^\ell \delta_{o_j, B}
%}
%The Kronecker $\delta$ in \eqref{L2} counts only texts with the right number of each word. We have
%\eq{
%\delta_{C(o), \ell \tilde p} &= \prod_{B=1}^T \delta_{C(o)_B, \ell \tilde p_B} \\
%&= \int \frac{d^T \lambda}{2\pi} e^{i \sum_B \lambda_B ( C(o)_B - \ell \tilde p_B) } \\
%&\equiv \int \D \lambda \; e^{i \sum_B \lambda_B ( \sum_{j=1}^\ell \delta_{o_j, B} - \ell \tilde p_B) } \\
%}
%The energy depends on the words through the term
%\eq{
%\sum_{a,B} \rho_{aB} \log O_{aB} = \sum_{a,B} \left[ \sum_{j=1}^\ell \delta_{o_j,B} \delta_{\sigma_j,a} \right] \log O_{aB} 
%}

\begin{figure*}[th!]
\centering
%\begin{subfigure}[b]{0.49\textwidth}\centering
\includegraphics[width=\textwidth]{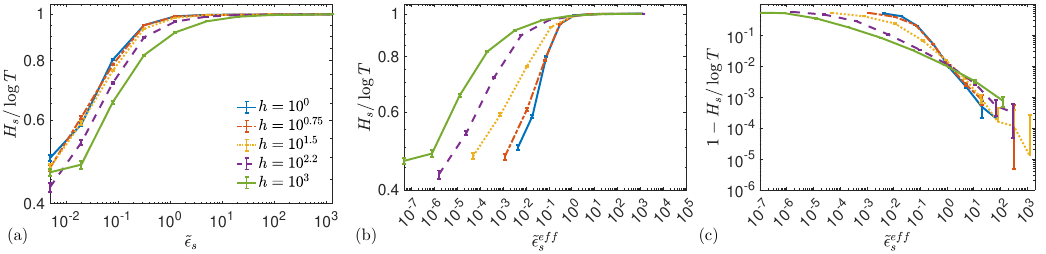}
\caption{The RLM transition is robust to the addition of a Zipfian surface bias. Curves are shown at $T=100$, $\tilde\epsilon_d \approx 0.03$, and indicated values of bias strength $h$; (a) the surface entropy versus $\tilde\epsilon_s$; bias increases from left to right; (b) the surface entropy versus an effective $\tilde\epsilon_s^{eff}(\tilde\epsilon_s,h)$ (see text). The onset of nontrivial surface entropy occurs at approximately $\tilde\epsilon_s^{eff}\approx 1$, but its development is weaker at larger biases. In (c) the same data from (b) is shown as an approach to the trivial value $H_s \to \log T$, valid as $\tilde\epsilon_s \to \infty$. All curves intersect approximately at $\tilde\epsilon_s\approx 1$. }
\label{fig3}
\end{figure*}

She thus obtains an estimate $\tilde \rho_{aB}$ of the joint word \& POS frequency, $\rho_{aB}$. Then she maximizes the likelihood of $\tilde \rho$, 
\eq{
\LL(M,O; \tilde \rho) & = \log \PP( \tilde \rho | M, O) \\
& = \log \sum_{\{\TT, \sigma, o\}} \delta_{\rho(o,\sigma), \tilde \rho } \PP(\TT,\sigma,o | M,O) \label{L2}
}
where $\rho(o,\sigma)_{aB}$ is the count of word $B$ and POS tag $a$ in the text of total length $\ell$, i.e.
\eq{
\rho(o,\sigma)_{aB} = \sum_{j=1}^\ell \delta_{o_j, B} \delta_{\sigma_j, a}
}
The Kronecker $\delta$ in \eqref{L2} counts only texts with the right number of each word and POS tag. We have
\eq{
\delta_{\rho(o,\sigma), \ell \tilde \rho} &= \prod_{a=1}^N \prod_{B=1}^T \delta_{\rho(o,\sigma)_{aB}, \tilde\rho_{aB}} \\
&= \prod_{a,B}\int_0^{2\pi} \frac{d \lambda_{aB}}{2\pi} e^{i \lambda_{a,B} ( \rho(o,\sigma)_{aB} - \tilde \rho_{aB}) } \\
&\equiv \int \D \lambda \; e^{i \sum_{a,B} \lambda_{a,B} ( \sum_{j=1}^\ell \delta_{o_j, B} \delta_{\sigma_j, a} - \tilde \rho_{aB}) } 
}
The energy depends on the words through the term
\eq{
\sum_{a,B} \rho_{aB} \log O_{aB} = \sum_{a,B} \left[ \sum_{j=1}^\ell \delta_{o_j,B} \delta_{\sigma_j,a} \right] \log O_{aB} 
}
which has the same dependence on the text and POS tags. So we can write
\eq{
\LL(M,O; \tilde \rho) = \log \frac{1}{Z} \sum_{\{\TT, \sigma, o\}}  \int \D \lambda \; e^{-i \lambda : \tilde \rho} e^{-E(\TT,\sigma,o | M, O(\lambda))} 
}
where
\eq{
\log O(\lambda)_{aB} = \log O_{aB} + i \lambda_{aB}
}
is a shifted surface grammar (in the complex plane). Note however that when a saddle point is attained (as will be the case for large texts), $i\lambda$ will be real, so that the grammar is real-valued as it must be. 

Finally $\LL$ becomes
\eq{ \label{Lfinal}
\LL(M,O; \tilde \rho) = \log \frac{1}{Z} \int \D \lambda \; e^{-i \lambda : \tilde \rho} Z(M,O(\lambda)) ,
}
so the likelihood depends on a shifted grammar. If we can evaluate this then we can derive the maximum-likelihood learning strategy, under the given assumptions. 

However $\LL$ is evaluated, the natural learning strategy on the grammars is simply to go in the gradient of increasing likelihood:
\eq{
\frac{d M_{abc}}{dt} & = k \frac{\p \LL}{\p M_{abc}} \\
\frac{d O_{aB}}{dt} & = k \frac{\p \LL}{\p O_{aB}} ,
}
where $k$ is the learning rate. 

Roughly speaking, $\LL$ is a difference of (minus) free energies: that of the RLM in the presence of a biased grammar (to match the observed $\tilde\rho$), but subtracting off the original RLM free energy. Thus the simple picture of \cite{DeGiuli19} is slightly modified: the learning scenario can be viewed as a free energy descent, but only along the directions that lower the free energy coupled to the correct biased grammar; if a change in the grammar equally affects $Z(M,O(\lambda))$ and $Z(M,O)$, then it will cancel out of $\LL$.

Let us try to understand \eqref{Lfinal} better. It involves the RLM partition function for a biased $O$ matrix. Note in general that 
\eq{
\frac{\p \log Z}{\p \log O_{aB}} = \frac{1}{Z} \sum_{\{\TT, \sigma, o\}} \rho_{aB} \; e^{-E} = \langle \rho_{aB} \rangle .
}
Now it is known that natural languages exhibit Zipf's law: the probability of a word decreases as a power law of its rank. Thus $\rho_{aB}$ will exhibit such behavior, and by this computation, so should the dependence of $\log Z$ on $\log O_{aB}$. Thus to understand $Z(M,O(\lambda))$ we should simulate the RLM in the presence of a bias $i\lambda$, which we take to have a Zipfian form. We consider this next.

\begin{figure*}[th!]
\centering
%\begin{subfigure}[b]{0.49\textwidth}\centering
\includegraphics[width=.8\textwidth]{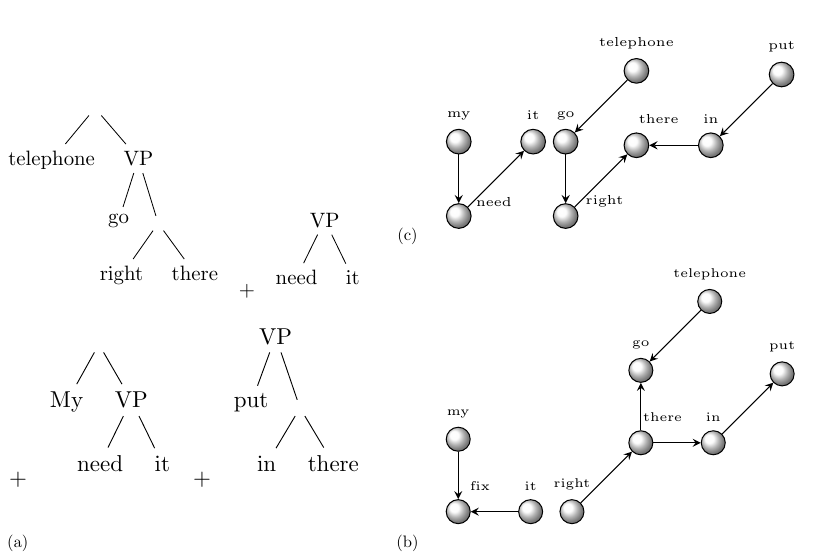}
\caption{Example syntax forest (a), dependency graph (b), and directed sentence graph (c) obtained from human data. Note that the word `fix' appeared in the dependency graph of \cite{Corominas-Murtra09} but not in the syntax tree shown therein.  }\label{figtrees}
\end{figure*}

\section{RLM with a bias}
%It is intuitively clear that the child's environment will favor certain 

The learning scenario motivates considering the RLM with a bias in the surface grammar. Consider
\eq{
\log O'_{aB} = \log O_{aB} + h_{aB}
}
where $h$ is the bias, and $O$ is given the distribution from the RLM. Then $O'$ has the distribution
\eqs{
\PP(O') & \propto \prod_{a,B} \frac{1}{O_{aB}} e^{-\tilde\epsilon_s \sum_{a,B} \log^2 (O'_{aB} e^{-h_{aB}}/\overline{O} ) } \notag \\
& \propto \prod_{a,B} \frac{1}{O'_{aB}} e^{-\tilde\epsilon_s \sum_{a,B} \log^2 (O'_{aB} /\overline{O} ) } e^{\tilde\epsilon_s \sum_{a,B} h_{aB} \log O'_{aB}/\overline{O} }
}
In order to disentangle the effect of the bias from that of $\epsilon_s$, we take $h_{aB} \propto 1/\tilde \epsilon_s$. As a Zipfian form, we consider
\eq{
h_{aB} = \frac{h}{\tilde\epsilon_s} \frac{1}{B} , 
}
where we arbitrarily order the words in decreasing rank. The scalar $h$ is the bias strength.

We simulated the RLM with Zipfian bias and a variety of field strengths, for $T=100$ and $\tilde\epsilon_d\approx0.03$. The resulting $H_s$ is shown in Fig.\ref{fig4}. The RLM transition is present in all cases, but its position depends on the bias strength $h$. A larger bias causes the transition to occur earlier (at higher $\tilde\epsilon_s$). This is intuitively clear, as the RLM transition was shown to induce the breaking of symmetries among symbols \cite{DeGiuli19}; since the bias breaks this symmetry explicitly, the transition occurs at higher $\tilde\epsilon_s$.

Inspecting Fig.\ref{fig4}a, it appears as though the data for different magnitudes of $h$ (`bias strengths') should collapse with some rescaled version of $\epsilon_s$. This suggests that a simple model may capture the  dependence on the bias. The transition discussed in \cite{DeGiuli19,DeGiuli19a,De-Giuli22} is controlled by the heterogeneity of the grammar, measured in the original model by \eqref{s1}, which satisfy \eqref{s2}. Thus we can see how $\overline{s_s}$ is renormalized by the bias. We evaluate
\eq{
\overline{s_s(h)} & \equiv \frac{1}{NT} \sum_{a,B} \overline { \log^2\left[\frac{O_{aB}e^{h_{aB}}}{\overline{O}} \right] } \\
& = \frac{1}{NT} \sum_{a,B} \int \frac{dO_{aB}}{O_{aB}\sqrt{\pi/\tilde\epsilon_s}} \log^2\left[\frac{O_{aB}e^{h_{aB}}}{\overline{O}} \right] e^{-\tilde\epsilon_s \log^2\left[\frac{O_{aB}}{\overline{O}} \right] } \notag\\
%& = \frac{1}{NT} \sum_{a,B} \int \frac{do_{aB}}{\sqrt{\pi/\tilde\epsilon_s}} \big(o_{aB}+h_{aB} - \log \overline{O} \big)^2 e^{-\tilde\epsilon_s \big( o_{aB} - \log \overline{O} \big)^2 }\notag\\
& = \frac{1}{NT} \sum_{a,B} \int \frac{do_{aB}}{\sqrt{\pi/\tilde\epsilon_s}} \big(o_{aB}+h_{aB}\big)^2 e^{-\tilde\epsilon_s o_{aB}^2  }\notag\\
& = \frac{1}{NT} \sum_{a,B} \left[ \frac{1}{2\tilde \epsilon_s} + (h_{aB})^2 \right] \notag\\
& = \frac{1}{2\tilde \epsilon_s} + \overline { h^2 } 
}
We can define a renormalized $\tilde \epsilon_s$ by
\eq{
\frac{1}{2 \tilde \epsilon_s^{eff}} =  \frac{1}{2\tilde \epsilon_s} + \overline { h^2 } 
}
As shown in Figs. 3b, this approximately collapses the initial decay of $H_s$ from its trivial value. Looking at this initial decay on a logarithmic scale (Fig 3c), all curves appear to cross at a common point $\tilde\epsilon_s^{eff}\approx 1$.

We also simulated the RLM with a staggered field of the form $h_{aB} = h/\tilde\epsilon_s \times g_B$, where $g_B$ takes only three values $1, \sqrt{1/T}, $and $1/T$, for the first, second, and third third of the symbols, respectively. The form and scaling is chosen to have a similar overall amplitude as the Zipfian bias. We found that for the same values of $h$ as above, there was no effect of the bias on $H_s$. We return to this later.

\section{Comparison with human data}
\begin{figure*}[th!]
\centering
%\begin{subfigure}[b]{0.49\textwidth}\centering
\includegraphics[width=.7\textwidth]{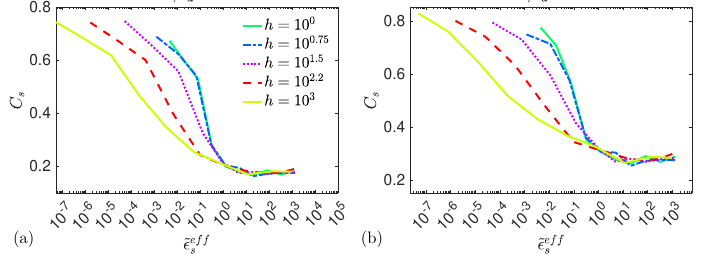}
\caption{The clustering coefficient of word graphs increases at the RLM transition, for $T=100$ and indicated Zipfian biases, with strength $h$. Both (a) directed and (b) undirected graphs show a similar increase of clustering around the transition point  $\tilde\epsilon_s^{eff}\approx 1$. In both plots, the bias increases from right to left at the top. }\label{fig4}
\end{figure*}
\begin{figure*}[th!]
\centering
%\begin{subfigure}[b]{0.49\textwidth}\centering
\includegraphics[width=.7\textwidth]{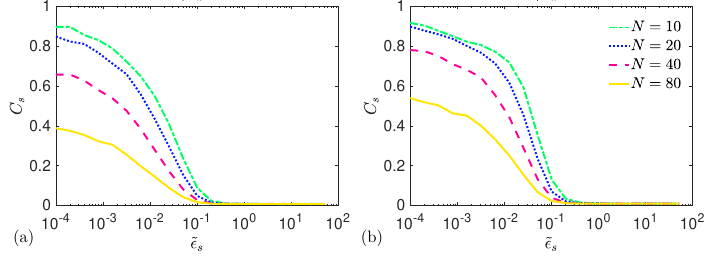}
\caption{The clustering coefficient of word graphs increases at the RLM transition, for $T=1000$ and indicated $N$. Both (a) directed and (b) undirected graphs show a similar increase of clustering around  $\tilde\epsilon_s\approx 0.1$. }\label{fig5}
\end{figure*}

How does the RLM compare to first language acquisition in children?

In previous work, syntactic networks were built from data of children's utterances between 22 and 32 months of age \cite{Corominas-Murtra09}, with data from the Peters corpora \cite{Bloom74,Bloom75}. The networks were built from dependency structures, with a mix of automated and manual procedures. These structures are graphs that connect observable symbols, related to but distinct from phrase structure trees. { Their aim is to represent, in a linear fashion, the dominant relationships between words; for example subject-verb, or modifier-head. In Ref.\cite{Corominas-Murtra09},} a variety of network-theoretic quantities showed a clear transition around 24 months of age; for example, both the word degree (the number of other words used with a given word) and the clustering coefficient (measuring the extent to which words are clustered) increase dramatically at this transition. { Quantitatively, the clustering is found to be less than 0.01 before age 22.5 months, and above 0.08 after 24 months. The maximal value shown is 0.2, at age 26.5 months. }

If the RLM is to apply to first language acquisition, then we should be able to see similar behaviors in these quantities, { in appropriate graphs constructed from syntax trees. However, the latter are not equivalent to the dependency graphs. In the setting of the RLM where words have no semantic meaning, there is no unambiguous way to create dependency structures. Therefore we build `sentence graphs' } as follows: we take the observed sentences $(o_1,o_2,\ldots,o_j,\ldots)$ and add a link to the graph from $B$ to $A$ if $o(j)=B,o(j+1)=A$, for some observable symbols $A$ and $B$ and index $j$. This directed graph includes many true dependency relations, but also spurious ones that would be absent in a more complete analysis. It gives a first approximation to the dependency graphs. 

{ To illustrate the similarities and differences between our graphs and those in \cite{Corominas-Murtra09}, in Fig.\ref{figtrees}a we reproduce the subset of syntax trees shown in \cite{Corominas-Murtra09}, along with their dependency graph in Fig.\ref{figtrees}b. In Fig.\ref{figtrees}c we show our directed sentence graph. One can see that the undirected structure of the graphs is very similar, while the direction of links is not always the same. For example, for the phrase ``telephone go right there'' the dependency graph identifies `go' as the head and points links towards it, while in our directed graph the links follow the final phrase ordering. As a result of this incomplete matching of the edge directions, we investigated both the directed graph described above, along with the undirected version where edges are not directed.}

{ In first language acquisition, both the size of the vocabulary and the manner in which the words are used changes as the child learns. For simplicity, in comparison with the RLM we will consider a situation where the vocabulary is fixed. This is motivated by the fact that, in the RLM, the position of the transition scales with $\epsilon_d \log N$ if controlled by $\epsilon_d$ and $N$, or $\epsilon_s \log T$ if controlled by $\epsilon_s$ and $T$: these show a  weak logarithmic dependence on the number of symbols/words, so that we expect the $\epsilon$'s to characterize the dominant changes during learning; future work could consider an explicit model for how $N$ and $T$ change during learning. Therefore in what follows we focus on the clustering coefficient and the degree distribution, both of which can be meaningfully compared regardless of $N$ and $T$. 

Initially, we consider the path $\gamma_3$ in Fig.\ref{fig_phase}. We confirmed that the results are very similar along path $\gamma_2$ (results not shown).
}

We investigated the clustering coefficient both for the directed graph, constructed as above, and the undirected graph constructed by adding the reverse links. The resulting clustering coefficients are shown in Fig.\ref{fig4} and Fig.\ref{fig5}. As the bias is varied, a clear increase is observed around $\tilde\epsilon_s^{eff}\approx 0.1$, consistent with the drop in sentence entropy. Similarly, as $N$ is varied the clustering also increases around the transition point. { Since a very similar behavior is observed for both our directed and undirected graphs, we expect that the match between this result and that found in \cite{Corominas-Murtra09} is not a coincidence.}

The linguistic interpretation of this behavior is interesting \cite{Corominas-Murtra09}:  the transition marks the point where the child begins to use functional items like $a$ or $the$ to connect many words. It thus represents the learning of a particular class of grammatical rules.

Ref. \cite{Corominas-Murtra09} also looked at the degree distribution of dependency graphs, finding that below the transition graphs were scale-free with $\PP(d_s)\sim 1/d_s^{1.3}$. { No information was given on the behavior of the distribution during learning. To compare with the degree distributions measured in \cite{Corominas-Murtra09}, we measured the degree distribution of our sentence graphs, shown in Fig.\ref{fig_deg}. We find that a power-law regime can be discerned, $\PP(d_s)\sim 1/d_s^{\gamma}$, but with an exponent that depends on $\tilde\epsilon_s$. In general, we find that the exponent decreases in magnitude as $\tilde\epsilon_s$ decreases. At $\tilde\epsilon_s= 10^{-2.2}$, the exponent matches what was observed in \cite{Corominas-Murtra09}, but we note that this result does not appear to be stable at lower $\tilde\epsilon_s$, where a hump develops at large degree. Moreover other corpora show various exponents: in Ref.\cite{Cancho04} texts from Czech, German, and Romanian show exponents $2.3,2.2$, and $2.2$, respectively \footnote{These are the exponents of undirected graphs; exponents for in-degree and out-degree graphs are similar.}. Therefore, both the human data and the RLM show scale-free behavior in the nontrivial regime. A more complete analysis of the human data over the course of learning would permit a more refined comparison. 

Finally, we also looked at the clustering coefficient across paths $\gamma_1$ and $\gamma_2$ in Fig.\ref{fig_phase} (data not shown). We find that along $\gamma_1$, $C_s$ is consistently high $\sim 0.7-0.9$, while along $\gamma_2$, the trajectory is very similar to that along $\gamma_3$ shown in Fig.\ref{fig5}. This supports that the first-language-acquisition learning curve does not take place at fixed small $\tilde\epsilon_s$. 

\begin{figure*}[th!]
\centering
%\begin{subfigure}[b]{0.49\textwidth}\centering
\includegraphics[width=\textwidth]{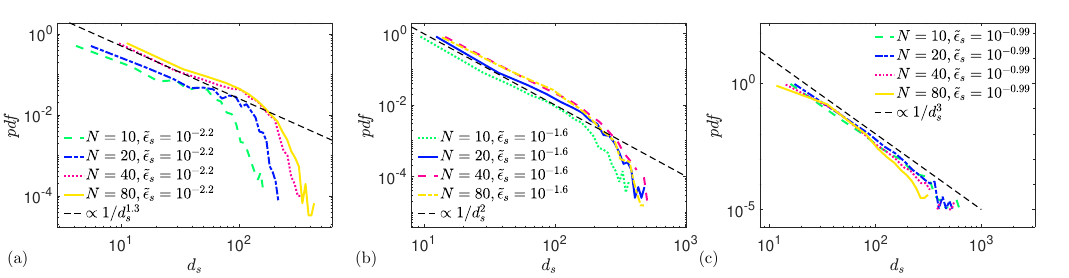}
\caption{Degree distribution of sentence graphs at indicated values of $N$ and (a) $\tilde\epsilon_s=10^{-2.2}$, (b) $\tilde\epsilon_s=10^{-1.6}$, (c) $\tilde\epsilon_s=10^{-0.99}$. In all cases an approximate power-law regime can be discerned. The shown lines have exponents $1.3, 2,$ and $3$, for (a,b,c), respectively.  }\label{fig_deg}
\end{figure*}

\begin{figure*}[th!]
\centering
%\begin{subfigure}[b]{0.49\textwidth}\centering
\includegraphics[width=\textwidth]{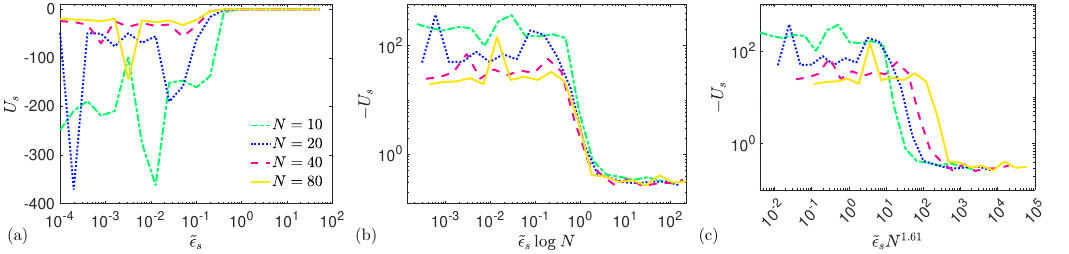}
\caption{Binder cumulant of observable word distribution, for $T=1000$ and indicated $N$. (a) This quantity begins to differ from 0 around $\tilde\epsilon_s\approx 1$, as expected. (b,c) On a logarithmic axis, the onset appears to collapse with a logarithmic factor of $N$, but not the power-law suggested in \cite{Nakaishi22}. }\label{fig_bin}
\end{figure*}

Overall, these results support that the RLM captures the initial onset of learning grammatical structure in first language acquisition. }

\section{Finite-size scaling}

True thermodynamic phase transitions only occur in the thermodynamic limit, because in a finite system, the partition function is an analytic function of control parameters. In the RLM, there are 2 distinct ways in which systems can be large: first, the sentence size $\ell$ gives the size of derivation structures, while $N$ and $T$ are the alphabet sizes, controlling the potential complexity of grammars. For this reason, in \cite{DeGiuli19} the senior author tuned the control parameters such that sentences were large (with a cutoff $\ell_{max} \sim 1000$), and moreover crucial observables were shown at various $N$. The existence of finite-size scaling in $N$ over an appreciable range from $N=10$ to $N=40$, and here up to $N=80$, shows that the basic phenomena of the RLM are not particular to small or large $N$. 

A recent work \cite{Nakaishi22} questioned whether the RLM shows a true thermodynamic phase transition. By a combination of analytic and numerical arguments, the authors argue that there is no phase transition at finite $\tilde\epsilon_d$ and finite $N$ in the RLM. However, as already shown in \cite{DeGiuli19}, to obtain satisfactory collapse of the data, quantities need to be collapsed with $\tilde\epsilon_d \log^\alpha N$, where $\alpha=1$ or $\alpha=2$ depending on the quantity considered. This is confirmed by theory that predicts $\alpha=1$, see for example \eqref{Hd} (after division by $\log N$ to compare with numerical results). 

Ref.\cite{Nakaishi22} measured in particular the Binder cumulant
\eq{
U = 1 - \frac{\langle (\pi_a-\langle \pi_a \rangle)^4 \rangle}{3 \langle (\pi_a-\langle \pi_a \rangle)^2 \rangle^2} ,
}
which is 0 if $\pi_a$ is Gaussian, and nonzero otherwise. Here $\pi_a$ is the empirical probability of observing hidden symbol $a$, related to the order parameter $Q_2$. Ref.\cite{Nakaishi22} found that $U$ has a dip at the transition, which becomes infinitely deep as $N\to \infty$, suggesting that the RLM becomes a true thermodynamic phase transition in this limit. Ref.\cite{Nakaishi22} suggest that the $\tilde\epsilon_d$ at which the minimum of $U$ is obtained goes to zero as $1/N^{1.61}$ but their fit is suspect: at the largest values of $N$ that they use (only $N=10$) the plot of $\log \tilde\epsilon_d$ versus $\log U$ has a distinct curvature, indicating that functional dependence on $N$ is not a power-law. It would indeed be very strange if $U$ did not collapse with $\tilde\epsilon_d \log^\alpha N$ as all other quantities do. The difference between $N^{1.61}$ and $\log^2 N$ in the range of small $N=1,\ldots,10$ considered by Ref.\cite{Nakaishi22} is slight. 

We measured the same quantity over an ensemble controlled by $\tilde\epsilon_d$ but found that the fluctuations in this quantity were huge, indicating that it is not self-averaging. Instead we found cleaner measurements of the Binder cumulant of $\pi_B$, the distribution of observable symbols, in the ensemble considered above, dependent upon $\tilde\epsilon_s$. As shown in Fig.\ref{fig_bin}, $U_s$ begins to differ from zero at the transition. On logarithmic axes, this onset appears to collapse with a logarithmic factor of $N$, but not the power law $N^{1.61}$ reported in Ref.\cite{Nakaishi22}; the much larger range of $N$ considered here allows us to distinguish these collapses much easier than would be possible in the range $N=1,\ldots,10$.  When the bias $h$ is varied, a similar behavior is observed (not shown).

It was mentioned in Ref.\cite{Nakaishi22} that the behavior of the Binder cumulant is similar to that observed in the 3D Heisenberg spin glass \cite{Imagawa02}. Thus, contrary to the title of Ref.\cite{Nakaishi22}, the results within actually support the existence of the RLM transition, in the limit $N\to\infty$, in appropriately rescaled variables. Since true thermodynamic phase transitions reside in universality classes, with a whole host of irrelevant variables, this further supports the robustness of the RLM as a simple model of syntax.

\section{Discussion}

{ The RLM encompasses all stochastic context-free grammars and, as such, is versatile. However, different applications may suggest different parameter ranges. This connects with ongoing discussion in linguistics on the relevant formalism to capture syntax of human languages. For example, in the classic rules-based approach of generative grammars, a child has to learn both the syntactic rules and the lexicon; in the RLM this means that their initial grammar would have large $\tilde\epsilon_d$ and large $\tilde\epsilon_s$. 

In the 1990's, Chomsky attempted to unify the CFGs of human languages by proposing in the Minimalist Program \cite{Chomsky14a} that their deep structure was essentially identical, captured by a Merge function that allows one to create tree-like derivation structures. Then variety among human languages would be captured by variety in the lexicon. More generally, this represented a shift from rules-based to constraint-based grammars. Although the associated merge grammars are, strictly speaking, different from CFGs, they maintain the core property of creating trees, and are similar to fixing a small $\tilde\epsilon_d$ so that deep structure is fixed. Then the learning problem would fix a small $\tilde\epsilon_d$ and allow the other parameters to vary, for example like path $\gamma_3$ in Fig.\ref{fig_phase}.

Along with the shift to constraint-based grammar, the Minimalist Program proposed that syntax requires an optimality computation, which was not specified in detail. This has been criticized as being unmotivated by core linguistic data \cite{Johnson97}, so it is not accepted as mainstream by linguists. For this reason, here we stay agnostic on the detailed description of learning and the relevant parameter ranges in the RLM, and focus on universal aspects.  

To learn a human language within the CFG framework, the Principles \& Parameters (P\&P) scenario for first language learning was proposed \cite{Chomsky93}. In it, }the task of learning syntax is reduced to the setting of a small number of discrete parameters, usually considered to be binary \cite{Shlonsky10}. Ongoing debate surrounds the detailed taxonomy of parameters and associated categorization of language, but regardless of these details, the scenario suggests that learning will occur in a series of discrete steps. Observables that quantify learning should then also show discrete steps. 

{ Meanwhile, connectionist models based on the physiology of the brain use continuous variables to learn \cite{McClelland81}. Debate on how people learn past-tense suggested the utility of stochastic rule-based models \cite{Pinker88}, like those considered in the RLM. While the early connectionist models gave poor performance, recent models do much better, without significant change in the underlying structure \cite{Kirov18}. Thus debate continues on the correct approach to learn syntax, with some calling for a more symbiotic approach between connectionism and generative grammar \cite{Pater19}.

The recent success of machine learning models at learning language has further ignited this debate \cite{Piantadosi23}. But while such models are an existence proof of the ability to learn language without significant constraints, they currently rely on a huge database to learn, and struggle with formal reasoning. Their connection to first-language acquisition in humans is thus unclear.

At variance with the P\&P approach, but more aligned with stochastic models and neural networks, 
human data analyzed in Ref.\cite{Corominas-Murtra09} as well as the RLM both suggest a single learning transition, with continuous (although in some cases quite abrupt) variation in observables. In the RLM this statement is robust to the inclusion of a bias, reflecting heterogeneity in the environment \footnote{One may wonder if the specific Zipfian bias considered above is itself too smooth to see a series of discrete transitions. To this end, we also tested a bias taking on only 3 values. Over the same range of bias strengths shown above, this bias did not have any effect on the sentence entropy.}. Thus, in all cases considered, the RLM transition is unimodal, matching that seen in human data.

These results suggest two possibilities. The first is that learning is truly a continuous process, in which 
%Of course, it may be that discrete transitions can be only be detected by sufficiently sensitive order parameters. But if, for the moment, we set aside this scenario, }and take seriously the observations of a continuous learning process, then it raises questions for first language acquisition. A continuous learning process suggests that 
}what is learned are weights (or probabilities) rather than discrete rules. Frequency effects are indeed ubiquitous in first language acquisition \cite{Ambridge15}, and there are proposals on how measured frequencies can be used to infer rules \cite{Yang17}. Moreover, the recent successes of machine learning in natural language processing \cite{Chang23} invariably use approaches with parameters that can be continuously tuned during the training process. Thus the notion of discrete syntactic parameters that are set during learning appears overly simplistic, and may fail to account for the diversity of human languages, as has been argued by linguists and psychologists, with vociferous debate \cite{Evans09}. Instead our results suggest that learning is continuous; after the RLM transition, the entropy of children's speech continuously decreases, and concominantly, the grammar becomes more and more certain. 
{

The second possibility is that discrete rules are indeed learned, but they are only detectible by sufficiently sensitive order parameters. Recent work on learning of {\it semantic} information showed a mechanism for discrete-like transitions hidden within a continuous process \cite{Saxe19}. Focusing on an input-output correlation matrix, it was found that singular values of this matrix are learned in a stepwise fashion; moreover when data is hierarchical, then these singular values are strongly graded, leading to distinct learning transitions. If this scenario also applies to learning of syntax, then it remains elusive in the data. }

\section{Conclusion}

The Random Language Model was introduced in \cite{DeGiuli19} as a simple model of language. We showed here that the RLM transition: (i) can be encountered by a change in properties of observable sentences; (ii) is robust to the inclusion of a bias, and (iii) is apparently a sharp thermodynamic transition as $N\to \infty$, in appropriately rescaled variables. A comparison with human data \cite{Corominas-Murtra09} supports that the RLM transition is equivalent to that experienced by most children in the age 22-26 months in the course of first language acquisition.

In future work, two avenues look promising: first, although limited by availability of quantitative data, more attempts to make a quantitative comparison with human data would be worthwhile; second, the astounding success of machine learning models to model natural language, and the lack of a theory to explain this, suggest that the RLM might shed light on this process. Indeed, the RLM captures several features of real-world data (long-range correlations, hierarchy, and combinatorial structure) that are missing from most physics models, and needed to understand modern deep neural networks \cite{Mezard23}.

Finally, the search for an analytical solution to the RLM is ongoing. A promising approach \cite{DeGiuli19a,De-Giuli22} represents syntax trees as Feynman diagrams for an appropriate field theory, but this falls short of a complete solution. The results of \cite{Nakaishi22}, as well as the results here, suggest that one should look for a solution in the idealized limit $N \to \infty$.

%\cite{Ambridge15} % ubiquity of frequency effects in FLA

%{\bf Simulations: } If this scenario is to work, then the restricted partition function $Z(M,O(\lambda))$ should show a learning transition; that is, a transition from high entropy to low entropy grammars, which we can check numerically. What form should we take for $\lambda$? In principle we need this for all $\lambda$, but for large texts and large vocabularies we expect that a saddle-point is attained. Now it is known that natural languages exhibit Zipf's law: the probability of a word decreases as a power law of its rank. Naively we expect that other fields, like $\lambda$, will also behave as power-laws. So we simulate the RLM with a biased $O$ in the form of Zipf's law.

{\bf Acknowledgments: } EDG is supported by NSERC Discovery Grant RGPIN-2020-04762.

%\bibliographystyle{abbrv}
%\bibliography{../Gravity,../Glasses}
\bibliography{../../Language,../../Glasses}

\end{document}